\documentclass[aps,twocolumn,showpacs]{revtex4}

\usepackage{graphicx}
\usepackage{longtable} 

\begin{document}

%\preprint{}


\title{Kinetic theory of point vortex systems\\
from the Bogoliubov-Born-Green-Kirkwood-Yvon hierarchy}


\author{Mitsusada M. Sano}
%\thanks{}
\affiliation{
Graduate School of Human and Environmental Studies,\\
Kyoto University,\\
Sakyo, Kyoto 606-8501, Japan
}


\date{\today}

\begin{abstract}
Kinetic equations are derived from 
the Bogoliubov-Born-Green-Kirkwood-Yvon (BBGKY) hierarchy 
for point vortex systems in an infinite plane. 
As the level of approximation for the Landau equation, 
the collision term of the kinetic equation derived 
coincides with that by Chavanis 
({\it Phys.~Rev.~E} {\bf 64}, 026309 (2001)).
Furthermore, we derive a kinetic equation 
corresponding to the Balescu-Lenard equation for plasmas, 
using the theory of the Fredholm integral equation. 
For large $N$, 
this kinetic equation is reduced to the Landau equation above.  
\end{abstract}

% insert suggested PACS numbers in braces on next line\
%
% 05.20.Dd  Kinetic theory
% 47.32.Cc  Vortex dynamics 
% 05.20.Jj  Statistical mechanics of classical fluids 
% 05.20.-y  Classical statistical mechanics 
% 47.10.+g  General theory(Fluid dynamics) 
%\keywords{}

\maketitle

% -------------------------------------------------------------------
% -------------------------------------------------------------------
%--------------------------------------------------------------------
\section{Introduction} \label{sec1}
A point vortex system is a model of continuous two-dimensional 
(2D) inviscid fluid dynamics. 
In fact, a point vortex is an idealized vortex of real vortex 
in 2D fluids. However, the point vortex system carries 
some important properties of 
continuous 2D inviscid fluid dynamics\cite{Chorin,Newton_Book}. 
It is formulated as a Hamilton dynamical system\cite{Helmholtz,Kirchhoff}. 
Therefore, a standard statistical mechanical theory was developed 
by Onsager\cite{Onsager}. 
In his formulation, 
states of a point vortex system are classified into two categories: 
one is positive temperature states, and the other is 
negative temperature states. 
In the positive temperature states, 
point vortices are distributed in a scattered way. 
However, in the negative temperature states, 
huge vortices are formed in time-evolution. 
The negative temperature states are phenomena observed in earth's surface. 
That is a formation of huge vortices, such as hurricanes and typhoons. 
Onsager's ideas are recently reviewed by Eyink and Sreenivasan\cite{ES}. 
Since Onsager, researchers considered the equilibrium statistical 
mechanics of point vortex systems. 
The main purpose of their studies is to 
construct the equilibrium states of point vortex system
\cite{Kida,J-M,P-L,L-P, Seyler,SandON}. 
The Poisson-Boltzmann equation is used to obtain the equilibrium states. 
For some cases, the equilibrium states are analytically obtained\cite{TCL}.

The next stage of the statistical mechanics of point vortex systems 
focuses on nonequilibrium properties. 
Standard nonequilibrium statistical mechanics goes 
toward kinetic theory. 
Two authors have studied kinetic theory of point vortex system,  
i.e., by Marmanis\cite{Marmanis} and by Chavanis\cite{Chavanis}.
Marmanis considered a gas of binary pairs of positive and 
negative point vortices. 
Chavanis considered a gas of point vortices with the same circulation. 
In this paper, we are interested in Chavanis's results. 

The system considered by Chavanis is closely related 
to the system of non-neutral plasmas in the Malmberg trap
\cite{DF,FCFD,JD,SDFD,KISM,SKIS,SKSA}. 
The dynamics of the non-neutral plasma in the Malmberg trap 
is described by the 2D Euler equation in a circular domain, 
using the guiding center approximation. 
Thus, 
if we are interested in non-neutral plasma in the experimental situation, 
as an idealized model, we should consider a point vortex system
in a circular domain. 
From experiments on non-neutral plasmas, 
many interesting properties of vortex dynamics of the 2D Euler equation 
are now known\cite{DF,FCFD,JD,SDFD,KISM,SKIS,SKSA}: (1) Diocotron instability
(i.e., in other words, Kelvin-Helmholtz instability), 
(2) Violent relaxation, (3) Slow decay, 
(4) Vortex crystals, and (5) Merger of vortices. 
As a theoretical aspect, 
recently the slow decay was numerically analyzed
using the point vortex system\cite{KN1,KN2,SYYT}. 
Although there is a difference in the boundary condition, 
these properties introduced here are common 
in the point vortex systems in an infinite plane. 

Back to the point vortex systems in an infinite plane, 
Chavanis derived serveral kinetic equations 
for the point vortex system in an infinite plane, 
in which the point vortices have the same circulation,  
and estimated interesting physical quantities, 
like the diffusion coefficient and the drift term, 
by using his kinetic equations\cite{Chavanis,Chavanis_Lemou,Chavanis_BBGKY}. 
In this paper, we develop a kinetic theory through 
the Bogoliubov-Born-Green-Kirkwood-Yvon (BBGKY) hierarchy
\cite{Bogoliubov, Born_Green, Kirkwood, Yvon} 
(for the text, See \cite{Balescu}). 
We compare our theory with that by Chavanis. 
The main purposes of this paper is as follows:
(1) We derive a kinetic equation (i.e., the Landau equation), 
which is identical to that by Chavanis\cite{Chavanis,Chavanis_BBGKY}.  
(2) Futhermore, 
we derive a kinetic equation (i.e., the Balescu-Lenard equation), 
which includes more correlation, i.e., the collective effects. 
This is a new kinetic equation. 
(3) We show that for large $N$, this kinetic equation is reduced 
to the Landau equation. 

The organization of this paper is as follows. 
In \S~\ref{sec2}, the equations of motion for a point vortex system 
in an infinite plane are presented. 
In \S~\ref{sec3}, the treatment of the BBGKY hierarchy is shown. 
Two key equations for time-evolution of the distribution function and the 
correlation function are derived. Using these equations, 
we derive the Vlasov equation, the Landau equation 
and the Balescu-Lenard equation for point vortex system 
in an infinite plane. 
It is shown that for large $N$, the Balescu-Lenard equation 
is reduced to the Landau equation. 
In \S~\ref{sec8}, we summarize the results of this paper 
and give future problems. 
%---------------------------------------------------------------------------
\section{Equations of motion} \label{sec2}
Let us consider a point vortex system, which consists of 
$N$ point vortices with the same circulation $\gamma$ in an infinite plane. 
The hamiltonian of this system is given by
\begin{equation}
H = - \frac{\gamma^{2}}{4\pi} \sum_{i\neq j}^{N} 
\ln |{\bf r}_{i}-{\bf r}_{j}|, 
\end{equation}
where ${\bf r}_{i}=(x_{i},y_{i})$. 
The equations of motion are written by using the hamiltonian:
\begin{equation}
\gamma \frac{dx_{i}}{dt}  =  \frac{\partial H}{\partial y_{i}}, \; 
\gamma \frac{dy_{i}}{dt}  =  -\frac{\partial H}{\partial x_{i}}.
\end{equation}
Thus, the velocities of the $i$th point vortex 
in the $x$- and $y$-directions are given by 
\begin{eqnarray}
v_{i}^{(x)} & = &  \frac{\gamma}{2\pi} 
\sum_{j\neq i}^{N} 
\frac{({\bf r}_{i}-{\bf r}_{j})_{y}}{|{\bf r}_{i}-{\bf r}_{j}|^{2}}, \\
v_{i}^{(y)} & = & -\frac{\gamma}{2\pi} 
\sum_{j\neq i}^{N} 
\frac{({\bf r}_{i}-{\bf r}_{j})_{x}}{|{\bf r}_{i}-{\bf r}_{j}|^{2}}.
\end{eqnarray}
It is convenient to rewrite the velocity in the following form. 
\begin{equation}
{\bf v}_{i} = \sum_{j\neq i} {\bf v}(j \rightarrow i),
\end{equation}
where
\begin{eqnarray}
{\bf v}(j \rightarrow i) & = & -\frac{\gamma}{2\pi} 
{\bf z} \times \frac{{\bf r}_{i}-{\bf r}_{j}}{|{\bf r}_{i}-{\bf r}_{j}|^{2}}
\nonumber \\
& = & \frac{\gamma}{2\pi} \frac{1}{|{\bf r}_{i}-{\bf r}_{j}|^{2}} 
\; J \cdot {\bf r}_{ij},
\end{eqnarray}
$J$ is the $2\times 2$ symplectic matrix 
\begin{equation}
J = 
\left (
\begin{array}{cc}
0 & 1 \\
-1 & 0
\end{array}
\right ),
\end{equation}
${\bf z}$ is the unit vector along the z-axis, 
and 
${\bf r}_{ij}= {\bf r}_{i}-{\bf r}_{j}$. 
This system have several conserved quantities:  
(1) Energy, i.e., $H=E$.
(2) Angular impulse, i.e., $I = \gamma \sum_{i=1}^{N} |{\bf r}_{i}|^{2}$. 
(3) Linear impulse, i.e., ${\bf L}= \gamma \sum_{i=1}^{N} {\bf r}_{i}$.\\ 

Now we define the vorticity $\omega({\bf r};t)$,
\begin{equation}
\omega({\bf r};t) = \sum_{i=1}^{N} \gamma \delta({\bf r}-{\bf r}_{i}),
\end{equation}
and the stream function $\psi({\bf r};t)$,
\begin{equation}
\psi({\bf r};t) = - \frac{\gamma}{2\pi} \sum_{i=1}^{N}
\ln|{\bf r}-{\bf r}_{i}|.
\end{equation}
Using the stream function, the velocity of the $i$th point vortex is given by 
\begin{eqnarray}
{\bf v}_{i} & = & - {\bf z} \times \nabla \psi({\bf r}={\bf r}_{i};t) 
\nonumber \\
& = & J\cdot \nabla \psi({\bf r}={\bf r}_{i};t).
\end{eqnarray}
It is easily confirmed that 
the vorticity satisfies the two-dimensional Euler equation.
\begin{equation}
\frac{\partial\omega}{\partial t} + {\bf v} \cdot \nabla \omega = 0. 
\end{equation}
%--------------------------------------------------------------
\section{The BBGKY hierarchy}\label{sec3}
We define the distribution function of $N$-point vortex systems:
\begin{equation}
F=F({\bf r}_{1},{\bf r}_{2},\dots,{\bf r}_{N};t).
\end{equation}
The Liouvile equation for $N$-point vortex systems in an infinite 
plane is given by 
\begin{equation}
\frac{\partial F}{\partial t} = {\cal L} F,
\end{equation}
where 
\begin{equation}
{\cal L} = \sum_{i=1}^{N} \frac{1}{\gamma} 
\left ( 
\frac{\partial H}{\partial x_{i}} \frac{\partial }{\partial y_{i}}
-  
\frac{\partial H}{\partial y_{i}} \frac{\partial }{\partial x_{i}}
\right ).
\end{equation}
The Liouvillian ${\cal L}$ is rewritten as follows.
\begin{eqnarray}
{\cal L} & = & 
- \sum_{i=1}^{N} \sum_{j\neq i} \left ( 
v_{x}(j\rightarrow i) \frac{\partial }{\partial y_{i}}
+  
v_{y}(j\rightarrow i)\frac{\partial }{\partial x_{i}}
\right ) \nonumber \\
& = &
- \sum_{i<j}^{N}  {\bf v}(j\rightarrow i) \cdot \nabla_{ij},
\end{eqnarray}
where $\nabla_{i} = \frac{\partial}{\partial {\bf r}_{i}}$ 
and $\nabla_{ij}= \nabla_{i} - \nabla_{j}$.
Now we have used the fact that 
${\bf v}(j\rightarrow i)= -{\bf v}(i\rightarrow j)$.
We define ${\cal L}_{ij}$ as
\begin{eqnarray}
{\cal L}_{ij} & = & -{\bf v}(j\rightarrow i) \cdot \nabla_{ij} \nonumber \\
& = & -{\bf v}(j\rightarrow i) \cdot \nabla_{i} 
-{\bf v}(i\rightarrow j) \cdot \nabla_{j}. 
\label{eq:Lij}
\end{eqnarray}
Thus, the Liouvillian becomes
\begin{equation}
{\cal L} = \sum_{i<j}^{N} {\cal L}_{ij}. 
\end{equation}

We define the $s$-body reduced distribution function:
\begin{widetext}
\begin{eqnarray}
f_{s}({\bf r}_{1},\dots,{\bf r}_{s}) 
& = & \frac{N!}{(N-s)!}
\int d{\bf r}_{s+1}d{\bf r}_{s+2}\dots d{\bf r}_{N} \; 
F({\bf r}_{1},\dots,{\bf r}_{s},{\bf r}_{s+1},\dots,{\bf r}_{N}). 
\label{eq:RDF}
\end{eqnarray}
\end{widetext}
Carrying out a usual manipulation of the BBGKY hierarchy 
(See \cite{Balescu} in detail), 
we obtain the following time-evolution equations 
for the $s$-body reduced distribution 
function $f_{s}({\bf r}_{1},\dots,{\bf r}_{s})$:
\begin{equation}
\partial_{t} f_{0} = 0,
\end{equation} 
and
\begin{eqnarray}
& & \partial_{t}f_{s}({\bf r}_{1},\dots,{\bf r}_{s}) \nonumber \\
& = & 
\sum_{i<j} {\cal L}_{ij} f_{s}({\bf r}_{1},\dots, {\bf r}_{s}) \nonumber \\
& & + \sum_{i=1}^{s} \int d{\bf r}_{s+1} \; {\cal L}_{i,s+1} 
f_{s+1}({\bf r}_{1},\dots,{\bf r}_{s+1}).
\label{eq:BBGKY}
\end{eqnarray}
This is the BBGKY hierarchy for the $s$-body reduced distribution functions.
The time-evolution of the $s$-body reduced distribution function 
is determined by the $s$-body reduced distribution function and 
the $(s+1)$-body reduced distribution function. 

For $s=1$, we have
\begin{equation}
\partial_{t} f_{1}({\bf r}_{1}) = \int d{\bf r}_{2} \; 
{\cal L}_{12} f_{2}({\bf r}_{1},{\bf r}_{2}).
\label{eq:1-body_RDF}
\end{equation}
Now we use the $2$-body and $3$-body correlation functions: 
$g_{2}({\bf r}_{1},{\bf r}_{2})$ and 
$g_{3}({\bf r}_{1},{\bf r}_{2},{\bf r}_{3})$. 
The $2$- and $3$-body correlation functions are related to 
the $2$- and $3$-body reduced distribution functions as follows:
\begin{eqnarray}
f_{2}({\bf r}_{1},{\bf r}_{2}) & = & 
f_{1}({\bf r}_{1}) f_{1}({\bf r}_{2}) + g_{2}({\bf r}_{1},{\bf r}_{2}), 
\label{eq:2-body_RDF}\\
f_{3}({\bf r}_{1},{\bf r}_{2},{\bf r}_{3}) & = & 
f_{1}({\bf r}_{1}) f_{1}({\bf r}_{2}) f_{1}({\bf r}_{3}) 
+ f_{1}({\bf r}_{1}) g_{2}({\bf r}_{2},{\bf r}_{3}) \nonumber \\
& & + f_{1}({\bf r}_{2}) g_{2}({\bf r}_{1},{\bf r}_{3}) 
+ f_{1}({\bf r}_{3}) g_{2}({\bf r}_{1},{\bf r}_{2}) 
\nonumber \\
& & + g_{3}({\bf r}_{1},{\bf r}_{2},{\bf r}_{3}).
\label{eq:3-body_RDF}
\end{eqnarray}
Thus, the correlation function describes the deviation 
from the product of the reduced distribution functions.
Inserting eq.~(\ref{eq:2-body_RDF}) into eq.~(\ref{eq:1-body_RDF}), 
we obtain 
\begin{equation}
\partial_{t}f_{1}({\bf r}_{1}) = \int d{\bf r}_{2} \; 
\left ( {\cal L}_{12} f_{1}({\bf r}_{1})f_{1}({\bf r}_{2})
+ {\cal L}_{12} g_{2}({\bf r}_{1},{\bf r}_{2})
\right ).
\label{eq:starting-point}
\end{equation}
Inserting eq.~(\ref{eq:2-body_RDF}) and eq.~(\ref{eq:3-body_RDF}) 
into eq.~(\ref{eq:BBGKY}), 
the time-evolution equations of $f_{2}({\bf r}_{1},{\bf r}_{2})$ 
is obtained: 
\begin{eqnarray}
& & \partial_{t}f_{2}({\bf r}_{1},{\bf r}_{2}) \nonumber \\
& = & 
{\cal L}_{12} [f_{1}({\bf r}_{1})f_{1}({\bf r}_{2}) 
+ g_{2}({\bf r}_{1},{\bf r}_{2})] \nonumber \\
& & 
+ \int d{\bf r}_{3} \; \{ {\cal L}_{13} 
[f_{1}({\bf r}_{1})f_{1}({\bf r}_{2})f_{1}({\bf r}_{3}) \nonumber \\
& & + f_{1}({\bf r}_{1})g_{2}({\bf r}_{2},{\bf r}_{3}) 
+ f_{1}({\bf r}_{2})g_{2}({\bf r}_{1},{\bf r}_{3}) \nonumber \\
& & + f_{1}({\bf r}_{3})g_{2}({\bf r}_{1},{\bf r}_{2}) 
+ g_{3}({\bf r}_{1},{\bf r}_{2},{\bf r}_{3}) ] + (1 \Leftrightarrow 2) \}.
\label{eq:tevol_2-body_RDF}
\end{eqnarray}
Similarly, using eq.~(\ref{eq:starting-point}), 
inserting eq.~(\ref{eq:2-body_RDF}) into eq.~(\ref{eq:tevol_2-body_RDF}), 
and manipulating the resulting equations, 
the time-evolution equation of $g_{2}({\bf r}_{1},{\bf r}_{2})$ 
is obtained: 
\begin{eqnarray}
& & \partial_{t}g_{2}({\bf r}_{1},{\bf r}_{2}) \nonumber \\
& = & 
{\cal L}_{12} f_{1}({\bf r}_{1})f_{1}({\bf r}_{2}) + 
{\cal L}_{12} g_{2}({\bf r}_{1},{\bf r}_{2}) \nonumber \\
& &
+ \int d{\bf r}_{3} \; 
\{ {\cal L}_{13} f_{1}({\bf r}_{1})g_{2}({\bf r}_{2},{\bf r}_{3}) + 
{\cal L}_{23} f_{1}({\bf r}_{2})g_{2}({\bf r}_{1},{\bf r}_{3}) \nonumber \\
& & + ({\cal L}_{13}+{\cal L}_{23}) 
[ f_{1}({\bf r}_{3})g_{2}({\bf r}_{1},{\bf r}_{2}) + 
g_{3}({\bf r}_{1},{\bf r}_{2},{\bf r}_{3}) ] \}.
\label{eq:tevol_2-body_CF}
\end{eqnarray}
Equation (\ref{eq:tevol_2-body_CF}) is slightly different 
from the recent result by Chavanis\cite{Chavanis_BBGKY}, 
who also developed a BBGKY hierarchy for the point vortex gas. 
The reason for these differences is unknown. 

Following \cite{Chavanis_BBGKY}, we shall close the hiearchy of BBGKY 
equations by considering an expansion in powers of $1/N$ 
for $N\rightarrow +\infty$. 
In the large $N$-limit, i.e., the hydrodynamic limit, 
we should preserve the total circulation $\Gamma$. 
Thus, the circulation $\gamma$ is $\gamma = \Gamma/N$, 
where $\Gamma = \mbox{const}$. 
The order estimate of each function is 
$f\sim 1,\; g_{2} \sim 1/N,\; \gamma\sim 1/N$, 
and ${\cal L}_{ij} \sim 1/N$. 
Thus, in eq.~(\ref{eq:tevol_2-body_CF}), 
the first term is $\sim 1/N$. 
The second term is $\sim 1/N^{2}$. 
The integral part is $\sim 1/N$,  
since the integration over ${\bf r}_{3}$ gives a $N$-factor. 
The term including the term $g_{3}$ is $\sim 1/N^{2}$. 
In the following treatment, the function $g_{3}$ is omitted, 
since we cut the correlation, i.e., truncate a chain of the BBGKY hierarchy. 
%--------------------------------------------------------------
\subsection{Vlasov equation}\label{sec3-1}
A mean field approximation is performed by neglecting 
the term of the correlation function in eq.~(\ref{eq:starting-point}):
\begin{equation}
\partial_{t}f({\bf r}_{1}) = \int d{\bf r}_{2} \; {\cal L}_{12} 
f({\bf r}_{1})f({\bf r}_{2}).
\end{equation} 
Using eq.~(\ref{eq:Lij}), we obtain
\begin{equation}
\frac{\partial f({\bf r}_{1})}{\partial t} + 
\langle {\bf v}_{1} \rangle \cdot \nabla_{1} f({\bf r}_{1}) = 0,
\label{eq:Vlasov}
\end{equation}
where
\begin{equation}
\langle {\bf v}_{1} \rangle = 
\int d{\bf r}_{2} \;  f({\bf r}_{2}) (v_{x}(2\rightarrow 1) {\bf i} + 
v_{y}(2\rightarrow 1) {\bf j} ).
\end{equation}
This is a mean field equation for the point vortex system 
in an infinite plane. It is analogous to the Vlasov equation 
in plasma physics and in stellar dynamics. 
We should note that this equation, eq.~(\ref{eq:Vlasov}), 
is nothing but the 2D Euler equation. 
%--------------------------------------------------------------
\subsection{Landau equation}\label{sec3-2}
The next higher order approximation is started 
with eq.~(\ref{eq:starting-point}) preserving the correlation function:
\begin{equation}
\partial_{t}f({\bf r}_{1}) = \int d{\bf r}_{2} \; 
\left ( {\cal L}_{12} f({\bf r}_{1})f({\bf r}_{2})
+ {\cal L}_{12} g_{2}({\bf r}_{1},{\bf r}_{2})
\right ).
\label{eq:starting-point_Landau}
\end{equation}
The right hand side of eq.~(\ref{eq:starting-point_Landau}) 
is upto the order $1/N^{2}$. 
For the correlation function $g_{2}({\bf r}_{1},{\bf r}_{2})$, 
we approximate eq.~(\ref{eq:tevol_2-body_CF}) upto the order $1/N$. 
\begin{eqnarray}
& & \partial_{t}g_{2}({\bf r}_{1},{\bf r}_{2})  \nonumber \\
& = & {\cal L}_{12} f({\bf r}_{1})f({\bf r}_{2}) \nonumber \\
& & + \int d{\bf r}_{3} ( {\cal L}_{13} + {\cal L}_{23} )
f({\bf r}_{3}) g_{2}({\bf r}_{1},{\bf r}_{2}) \nonumber \\
& = & {\cal L}_{12} f({\bf r}_{1})f({\bf r}_{2}) + \nonumber \\
& & (- \langle {\bf v}_{1} \rangle \cdot \nabla_{1} 
- \langle {\bf v}_{2} \rangle \cdot \nabla_{2} )
g_{2}({\bf r}_{1},{\bf r}_{2}).
\label{eq:starting-point_CF}
\end{eqnarray}
Therefore, the correlation function is advected 
by $\langle {\bf v}_{1} \rangle$ and $\langle {\bf v}_{2} \rangle$. 
Equation (\ref{eq:starting-point_CF}) is formally solved as 
\begin{eqnarray}
& & g_{2}({\bf r}_{1},{\bf r}_{2};t) \nonumber \\
& = & 
U_{12}(t)g_{2}({\bf r}_{1},{\bf r}_{2};0) \nonumber \\
& & + \int_{0}^{t}d\tau \; U_{12}(\tau) 
{\cal L}_{12} f({\bf r}_{1};t-\tau)f({\bf r}_{2};t-\tau), 
\label{eq:solution_CF}
\end{eqnarray}
where
\begin{eqnarray}
U_{12}(\tau) 
& = & 
\exp \left [ 
- \int_{0}^{\tau} dt' \langle {\bf v}_{1} \rangle \cdot \nabla_{1} 
- \int_{0}^{\tau} dt' \langle {\bf v}_{2} \rangle \cdot \nabla_{2} \right ]. 
\end{eqnarray}
Inserting eq.~(\ref{eq:solution_CF}) 
into eq.~(\ref{eq:starting-point_Landau}), 
we obtain
\begin{widetext}
\begin{eqnarray}
\partial_{t}f({\bf r}_{1})
& = & \int d{\bf r}_{2}\;  
{\cal L}_{12} f({\bf r}_{1})f({\bf r}_{2})
+ \int d{\bf r}_{2} \; {\cal L}_{12} U_{12}(t)g_{2}({\bf r}_{1},{\bf r}_{2};0) 
\nonumber \\
& & 
+ \int d{\bf r}_{2} \; \int_{0}^{t} d\tau \; {\cal L}_{12} U_{12}(\tau) 
{\cal L}_{12} f({\bf r}_{1};t-\tau) f({\bf r}_{2};t-\tau). 
\label{eq:step1}
\end{eqnarray}
In the right hand side of eq.~(\ref{eq:step1}), 
the second term vanishes for large $t$ (i.e., the correlation decays.), 
thus we have 
\begin{eqnarray}
\partial_{t}f({\bf r}_{1})
& = & \int d{\bf r}_{2} \; 
{\cal L}_{12} f({\bf r}_{1})f({\bf r}_{2})
+ \int d{\bf r}_{2} \; \int_{0}^{t} d\tau \; {\cal L}_{12} U_{12}(\tau) 
{\cal L}_{12} f({\bf r}_{1};t-\tau) f({\bf r}_{2};t-\tau).
\end{eqnarray}
We have to evaluate the following term:
\begin{eqnarray}
{\cal K}_{\mbox{\scriptsize coll}}^{(L)}\{ff\} 
& =&  
\int d{\bf r}_{2} \; \int_{0}^{t} \; d\tau \; {\cal L}_{12} U_{12}(\tau) 
{\cal L}_{12} f({\bf r}_{1};t-\tau) f({\bf r}_{2};t-\tau).
\label{eq:generalized_keq_collision_term}
\end{eqnarray}
Using the approximation, 
\begin{equation}
f({\bf r}_{1};t-\tau) f({\bf r}_{2};t-\tau) 
\approx U_{12}(-\tau) f({\bf r}_{1};t) f({\bf r}_{2};t), 
\end{equation}
we obtain
\begin{eqnarray}
{\cal K}_{\mbox{\scriptsize coll}}^{(L)}\{ff\}
& \approx &
\int d{\bf r}_{2} \; \int_{0}^{t} \; d\tau \; {\cal L}_{12} U_{12}(\tau) 
{\cal L}_{12} U_{12}(-\tau)f({\bf r}_{1};t) f({\bf r}_{2};t) \nonumber \\
& = & 
\nabla_{1} \cdot \int d{\bf r}_{2} \; \int_{0}^{t} \; d\tau \;
{\bf v}_{1}(t) {\bf v}_{1}(t-\tau)
\nabla_{12}f({\bf r}_{1};t) f({\bf r}_{2};t).
\label{eq:Landau_coll1}
\end{eqnarray}
The kinetic equation obtained here is 
\begin{equation}
\frac{\partial f({\bf r}_{1})}{\partial t} 
+ \langle {\bf v}_{1} \rangle \cdot \nabla_{1} f({\bf r}_{1})
= \nabla_{1}\cdot \int d{\bf r}_{2} \; \int_{0}^{t} d\tau \; 
{\bf v}_{1}(t) {\bf v}_{1}(t-\tau)\; 
\cdot \nabla_{12}f({\bf r}_{1};t) f({\bf r}_{2};t),
\label{eq:Landau}
\end{equation}
where 
${\bf v}_{1}(t)$ is advected as 
${\bf v}_{1}(t-\tau)=U_{12}(\tau){\bf v}_{1}(t)U_{12}(-\tau)$ 
and 
${\bf r}_{i}(t-\tau)= {\bf r}_{i}(t) 
- \int_{0}^{\tau} dt'\; \langle {\bf v}_{i} \rangle 
({\bf r}_{i}(t-t'),t-t')$. 
This equation is analogous to the Landau equation 
in plasma physics and in stellar dynamics. 
This equation coincides with the result of \cite{Chavanis,Chavanis_BBGKY}.
As shown in \cite{Chavanis}, this equation conserves 
the angular impulse and the linear impulse. 
If we use the Markovianization, 
i.e., extending the time integral to infinity, 
we obtain
\begin{eqnarray}
{\cal K}_{\mbox{\scriptsize coll}}^{(L)}\{ff\}
& \approx &
\nabla_{1} \cdot \int d{\bf r}_{2} \; \int_{0}^{\infty} d\tau \; 
{\bf v}_{1}(t) {\bf v}_{1}(t-\tau) \cdot
\nabla_{12}f({\bf r}_{1};t) f({\bf r}_{2};t).
\label{eq:Landau_coll2}
\end{eqnarray}
\end{widetext}
However, it is not known whether the Markovianization is assured or not, 
since point vortex dynamics sometimes gives long-time tail, 
i.e., the strong correlation. 
In particular, in \cite{KN2}, 
it is shown that the diffusion process for the point vortex 
exhibits L\'{e}vy flight. 

Chavanis estimated the relaxation time $t_{\mbox{\scriptsize relax}}$ 
by using the estimate of the diffusion coefficient\cite{Chavanis} 
as $t_{\mbox{\scriptsize relax}} \sim N/(\ln N) t_{D}$,  
where the dynamical time is 
$t_{D}\sim \langle \omega \rangle^{-1} \sim R^{2}/\Gamma$, 
which is the time determined by the mean rotation time, 
and $R$ is the size of the vortex. 
His estimate of \cite{Chavanis} would be incorrect. 
In the kinetic theory, 
the $N$-dependence of $t_{\mbox{\scriptsize relax}}$ is determined 
by the $N$-dependence of the collision term, i.e., 
${\cal K}^{(L)}_{\mbox{\scriptsize coll}} \sim O(1/N)$. 
This gives $t_{\mbox{\scriptsize relax}} \sim N t_{D}$.
Recently Chavanis and Lemou used 
this estimate\cite{Chavanis_Lemou,Chavanis_BBGKY}. 
This estimate is consistent with the numerical result 
by Kawahara and Nakanishi for the system in a circular domain\cite{KN2}. 
%-------------------------------------------------------------------------
\subsection{Balescu-Lenard equation} \label{new_sec}
In this subsection, we derive a kinetic equation 
for point vortex systems in an infinite plane, 
which is analogous to the Balescu-Lenard equation in plasma physics. 
The starting point is the time-evolution equations of 
the one-body reduced distribution function $f({\bf r}_{1})$ and 
the two-body correlation function $g_{2}({\bf r}_{1},{\bf r}_{2})$. 
\begin{equation}
\partial_{t} f({\bf r}_1) = \int d{\bf r}_{2}\; 
( {\cal L}_{12} f({\bf r}_{1})f({\bf r}_{2}) 
+ {\cal L}_{12} g_{2}({\bf r}_{1},{\bf r}_{2}) ),
\end{equation}
and
\begin{eqnarray}
& & \partial_{t}g_{2}({\bf r}_{1}, {\bf r}_{2}) \nonumber \\
& = & {\cal L}_{12}f({\bf r}_{1})f({\bf r}_{2}) + 
{\cal L}_{12} g_{2}({\bf r}_{1},{\bf r}_{2}) \nonumber \\
& & + \int d{\bf r}_{3} \; 
\{ {\cal L}_{13} f({\bf r}_{1}) g_{2}({\bf r}_{2},{\bf r}_{3}) 
+ {\cal L}_{23} f({\bf r}_{2}) g_{2}({\bf r}_{1},{\bf r}_{3}) \nonumber \\
& & + ({\cal L}_{13} + {\cal L}_{23}) [ 
f({\bf r}_{3}) g_{2}({\bf r}_{1},{\bf r}_{2}) ] \}.
\label{eq:t_evol_2body_CF}
\end{eqnarray}
The two-body correlation function is formally solved as
\begin{widetext}
\begin{eqnarray}
g_{2}({\bf r}_{1}, {\bf r}_{2};t)
& = & \int_{0}^{t} d\tau \, U_{12}(\tau) \, 
\{ {\cal L}_{12}U_{12}(-\tau)f({\bf r}_{1})f({\bf r}_{2}) + 
{\cal L}_{12}U_{12}(-\tau)g_{2}({\bf r}_{1},{\bf r}_{2}) \nonumber \\
& & + \int d{\bf r}_{3} \; 
( {\cal L}_{13} U_{12}(-\tau)f({\bf r}_{1}) g_{2}({\bf r}_{2},{\bf r}_{3}) 
+ {\cal L}_{23} U_{12}(-\tau)f({\bf r}_{2}) g_{2}({\bf r}_{1},{\bf r}_{3}) ) 
\}
\label{eq:2body_CF1}
\end{eqnarray}
\end{widetext}
The kinetic equation is formally obtained as
\begin{equation}
\frac{\partial f_{1}}{\partial t} + 
\langle {\bf v}_{1} \rangle \cdot \nabla_{1}f_{1} = 
\int d{\bf r}_{2} \; {\cal L}_{12} g_{2}({\bf r}_{1},{\bf r}_{2};t).
\label{eq:kinetic_eq0}
\end{equation}
If, as done for plasma systems in \cite{Balescu}, 
we set 
\begin{eqnarray}
g_{2}({\bf r}_{1},{\bf r}_{2};t) & = & g_{2}({\bf r}_{1}-{\bf r}_{2};t) 
\nonumber \\
& =& \int d{\bf k} \exp [i{\bf k}\cdot ({\bf r}_{1}-{\bf r}_{2})]
\tilde{g}_{2}({\bf k};t),
\label{eq:g2_homogeneous}
\end{eqnarray}
the right hand side of eq.(\ref{eq:kinetic_eq0}), i.e., the collision term, 
vanishes. 
\begin{eqnarray}
& &\int d{\bf r}_{2} {\cal L}_{12} g_{2}({\bf r}_{1},{\bf r}_{2};t)
\nonumber \\
& = & 
\frac{1}{\gamma} \int d{\bf r}_{2} \; 
(\nabla_{1}V({\bf r}_{1}-{\bf r}_{2}))^{\top} \cdot J \cdot 
\nabla_{1} g_{2}({\bf r}_{1},{\bf r}_{2}) \nonumber \\
& = & 
\frac{1}{\gamma} \int d{\bf r}_{2}\;
\int d{\bf k} \; \exp[i{\bf k}\cdot({\bf r}_{1}-{\bf r}_{2})]  
\tilde{V}(k) (i{\bf k})^{\top} \cdot J \nabla_{1} \nonumber \\
& & 
\times \int d{\bf k}' \; \exp[i{\bf k}'\cdot({\bf r}_{1}-{\bf r}_{2})]  
\tilde{g}_{2}({\bf k}';t)\nonumber \\
& = & 
\frac{(2\pi)^{2}}{\gamma} \int d{\bf k} \; 
\tilde{V}(k) \; {\bf k}^{\top}\cdot J \cdot {\bf k}\; 
\tilde{g}_{2}(-{\bf k};t)\nonumber \\
& = & 0,
\end{eqnarray}
since ${\bf k}^{\top}\cdot J \cdot {\bf k}=0$. 
Here ``${\bf A}^{\top}$'' means the transpose of the vector ${\bf A}$. 
In eq.~(\ref{eq:g2_homogeneous}), the homogeneity is assumed. 
The above result shows that inhomogeneity is important 
for point vortex systems. 
To not make the collision term vanish, 
we change the definition of the Fourier transform of 
$g_{2}({\bf r}_{1},{\bf r}_{2};t)$. 
Therefore, we set
\begin{eqnarray}
& & g_{2}({\bf r}_{1},{\bf r}_{2};t)  \nonumber \\
& =& \int d{\bf k}_{1} \int d{\bf k}_{2} \; 
\exp [i{\bf k}_{1}\cdot {\bf r}_{1} + i{\bf k}_{2}\cdot{\bf r}_{2}]
\tilde{g}_{2}({\bf k}_{1},{\bf k}_{2};t).
\label{eq:g2_inhomogeneous}
\end{eqnarray}
${\cal L}_{12}$ can be rewritten in the form 
\begin{equation}
{\cal L}_{12} = \frac{1}{\gamma} 
(\nabla_{1} V({\bf r}_{1}-{\bf r}_{2}))^{\top}\cdot J \cdot \nabla_{12},
\end{equation}
where
\begin{equation}
V({\bf r}_{1}-{\bf r}_{2}) = - \frac{\gamma^{2}}{2\pi} 
\ln |{\bf r}_{1}-{\bf r}_{2}| .
\end{equation}
Now we consider the Fourier transform of the function $V({\bf r})$, 
where ${\bf r}={\bf r}_{1}-{\bf r}_{2}$:
\begin{eqnarray}
V({\bf r}) & = & \int d{\bf k} \; \tilde{V}({\bf k}) 
e^{i{\bf k} \cdot {\bf r}},
\nonumber \\
\tilde{V}({\bf k}) & = & \frac{1}{(2\pi)^{2}} 
\int d{\bf r} V({\bf r}) e^{-i{\bf k} \cdot {\bf r}}.
\end{eqnarray}
The Fourier transform of $V({\bf r})$ is evaluated as follows:
\begin{eqnarray}
& & \tilde{V}({\bf k})  \nonumber \\
& = & -\frac{\gamma^{2}}{(2\pi)^{3}} 
\int d{\bf r} \; \ln|{\bf r}| 
e^{-i{\bf k} \cdot {\bf r}} \nonumber \\
& = & 
-\frac{\gamma^{2}}{(2\pi)^{3}} 
\int_{0}^{\infty} r\,dr \; \int_{0}^{2\pi} d\theta \; 
\ln r e^{-ikr\cos(\theta)} \nonumber \\
& = & 
-\frac{\gamma^{2}}{(2\pi)^{2}} 
\int_{0}^{\infty} dr \; 
r\ln r J_{0}(kr) \nonumber\\
& = & 
-\frac{\gamma^{2}}{(2\pi)^{2}} \left \{ 
\left [\frac{r}{k}\ln r J_{1}(kr) \right ]_{0}^{\infty} 
- \frac{1}{k}\int_{0}^{\infty} dr \; J_{1}(kr) \right \} . 
\end{eqnarray}
We have to evaluate the following limit:
\begin{equation}
\lim_{r\rightarrow \infty} \frac{r}{k}\ln r J_{1}(kr)
\end{equation}
In fact, for large $r$, this function oscillates with amplifying 
its absolute value. Therefore, we suppose that 
the limiting value of this function is zero. 
Alternatively, we insert a convergence factor:
\begin{equation}
\lim_{\epsilon \rightarrow +0}
\int_{0}^{\infty} dr \; r\ln r J_{0}(kr) e^{-\epsilon r}.
\end{equation}
As a result, we have
\begin{equation}
\tilde{V}({\bf k}) = \frac{\gamma^{2}}{(2\pi)^{2}} \frac{1}{k^{2}}.
\end{equation}
The dependence of $\tilde{V}({\bf k})\sim 1/k^{2}$ is a typical behavior 
of Coulomb systems. 

\begin{widetext}
Now we evaluate each term in the right hand side of eq.(\ref{eq:2body_CF1}). 
\begin{eqnarray}
(A) 
& = & 
\int_{0}^{t} d\tau \; 
U_{12}(\tau) {\cal L}_{12} U_{12}(-\tau)f_{1}f_{2} \nonumber \\
& = & 
-\int_{0}^{t} d\tau \; {\bf v}_{1}(t-\tau) \cdot 
\nabla_{12}f_{1}f_{2} \nonumber \\
& = & 
\frac{1}{\gamma}
\int_{0}^{t} d\tau \int d{\bf k}  
\exp \left [ i{\bf k}\cdot({\bf r}_{1}-{\bf r}_{2}) 
- i{\bf k}\cdot \int^{\tau} dt' \langle {\bf v}_{1} \rangle 
+ i{\bf k}\cdot \int^{\tau} dt' \langle {\bf v}_{2} \rangle 
\right ] 
\tilde{V}(k) (i{\bf k})^{\top}\cdot J \cdot \nabla_{12}f_{1}f_{2}.
\end{eqnarray}

\begin{eqnarray}
(B)
& = & 
\int_{0}^{t} d\tau \; U_{12}(\tau) 
{\cal L}_{12} U_{12}(-\tau) g_{2}({\bf r}_{1},{\bf r}_{2};t)
\nonumber \\
& = & 
-\int_{0}^{t} d\tau \; {\bf v}_{1}(t) \cdot \nabla_{12}
g_{2}({\bf r}_{1},{\bf r}_{2};t) \nonumber \\
& = & 
\frac{1}{\gamma}
\int_{0}^{t} d\tau \int d{\bf k}  
\exp \left [ i{\bf k}\cdot({\bf r}_{1}-{\bf r}_{2}) 
- i{\bf k}\cdot \int^{\tau} dt' \langle {\bf v}_{1} \rangle 
+ i{\bf k}\cdot \int^{\tau} dt' \langle {\bf v}_{2} \rangle 
\right ] 
\tilde{V}(k) (i{\bf k})^{\top}\cdot J \cdot \nabla_{12}
g_{2}({\bf r}_{1},{\bf r}_{2};t).
\end{eqnarray}

\begin{eqnarray}
(C)
& = & 
\int_{0}^{t} d\tau \; U_{12}(\tau) 
\int d{\bf r}_{3} {\cal L}_{13} U_{123}(-\tau) 
f_{1} g_{2}({\bf r}_{2},{\bf r}_{3};t)
\nonumber \\
& = & 
\frac{1}{(2\pi)^{2}\gamma} 
\int d{\bf R} \int d{\bf R}' 
\int_{0}^{t} d\tau \int d{\bf k} \int d{\bf k}' \nonumber \\
& & 
\times \exp \left [i{\bf k}\cdot({\bf r}_{1}-{\bf R}') + 
i {\bf k}'\cdot ({\bf r}_{2}-{\bf R}) 
- i{\bf k}\cdot \int^{\tau} 
( \langle {\bf v}_{1} \rangle - \langle {\bf V}' \rangle )
dt' \right ]
\nonumber \\
& & 
\times 
\tilde{V}(k) ((i{\bf k})^{\top}\cdot J \cdot \nabla_{1}f_{1}) 
g_{2} ({\bf R}, {\bf R}';t ).
\end{eqnarray}

\begin{eqnarray}
(D)
& = & 
\int_{0}^{t} d\tau \; U_{12}(\tau) 
\int d{\bf r}_{3} {\cal L}_{23} U_{123}(-\tau) 
f_{2} g_{2}({\bf r}_{1},{\bf r}_{3};t)
\nonumber \\
& = & 
\frac{1}{(2\pi)^{2}\gamma} 
\int d{\bf R} \int d{\bf R}' 
\int_{0}^{t} d\tau \int d{\bf k} \int d{\bf k}' \nonumber \\
& & 
\times \exp \left [i{\bf k}'\cdot({\bf r}_{1}-{\bf R}) + 
i {\bf k}\cdot ({\bf r}_{2}-{\bf R}') 
- i{\bf k}\cdot \int^{\tau} 
( \langle {\bf v}_{2} \rangle - \langle {\bf V}' \rangle )
dt' \right ]
\nonumber \\
& & 
\times \tilde{V}(k) ((i{\bf k})^{\top}\cdot J \cdot \nabla_{2}f_{2}) 
g_{2} ({\bf R}, {\bf R}';t ).
\end{eqnarray}

The order estimate of these terms is as follows.
\begin{equation}
(A)\sim \frac{1}{N},\; 
(B)\sim \frac{1}{N^{2}},\; 
(C)\sim \frac{1}{N},\; 
(D)\sim \frac{1}{N}.
\end{equation}
Therefore, we can neglect the term of $(B)$. 

Then we obtain the following integral equation:
\begin{eqnarray}
g_{2}({\bf r}_{1},{\bf r}_{2};t)
& = &  
q({\bf r}_{1},{\bf r}_{2};t) 
+ \int d{\bf R} \int d{\bf R}' \; 
K(\{ {\bf r}_{1},{\bf r}_{2} \}, \{ {\bf R}, {\bf R}'\}) 
g_{2}({\bf R},{\bf R}';t),
\end{eqnarray}
where 
\begin{eqnarray}
q({\bf r}_{1},{\bf r}_{2};t)
& = & 
\frac{1}{\gamma}
\int_{0}^{t} d\tau \int d{\bf k}  
\exp \left [ i{\bf k}\cdot({\bf r}_{1}-{\bf r}_{2}) 
- i{\bf k}\cdot \int^{\tau} dt' \langle {\bf v}_{1} \rangle 
+ i{\bf k}\cdot \int^{\tau} dt' \langle {\bf v}_{2} \rangle 
\right ] \nonumber \\
& & 
\times \tilde{V}(k) (i{\bf k})^{\top}\cdot J \cdot \nabla_{12}f_{1}f_{2},
\end{eqnarray}
and 
\begin{eqnarray}
& & K(\{ {\bf r}_{1},{\bf r}_{2} \}, \{ {\bf R}, {\bf R}'\}) \nonumber \\
& = & 
\frac{1}{(2\pi)^{2}\gamma} 
\int_{0}^{t} d\tau \int d{\bf k} \int d{\bf k}' 
\exp \left [i{\bf k}\cdot({\bf r}_{1}-{\bf R}') + 
i {\bf k}'\cdot ({\bf r}_{2}-{\bf R}) 
- i{\bf k}\cdot \int^{\tau} 
( \langle {\bf v}_{1} \rangle - \langle {\bf V}' \rangle )
dt' \right ]
\nonumber \\
& & \times \tilde{V}(k) (i{\bf k})^{\top}\cdot J \cdot \nabla_{1}f_{1} 
\nonumber \\
&  & 
+ \frac{1}{(2\pi)^{2}\gamma} 
\int_{0}^{t} d\tau \int d{\bf k} \int d{\bf k}' 
\exp \left [i{\bf k}'\cdot({\bf r}_{1}-{\bf R}) + 
i {\bf k}\cdot ({\bf r}_{2}-{\bf R}') 
- i{\bf k}\cdot \int^{\tau} 
( \langle {\bf v}_{2} \rangle - \langle {\bf V}' \rangle )
dt' \right ]
\nonumber \\
& & \times \tilde{V}(k) (i{\bf k})^{\top}\cdot J \cdot \nabla_{2}f_{2}. 
\end{eqnarray}
\end{widetext}
The function $K(\{ {\bf r}_{1},{\bf r}_{2} \}, \{ {\bf R}, {\bf R}'\})$ 
is called the integral kernel of the integral equation. 
This integral equation takes the form of the Fredholm integral equation 
of the second kind. 
Thus, how to solve it is known\cite{CH}. 
Now for brevity, we set ${\bf x} = ({\bf r}_{1},{\bf r}_{2})$ and 
${\bf y} = ({\bf R},{\bf R}')$. 
The integral equation, which should be solved, is 
\begin{eqnarray}
g_{2}({\bf x};t)
& = &  
q({\bf x};t) 
+ \lambda \int d{\bf y} \; K({\bf x},{\bf y}) g_{2}({\bf y};t).
\label{eq:integral_eq_to_be_solved}
\end{eqnarray}
If the required conditions are satisfied, 
this integral equation is solved as 
\begin{eqnarray}
g_{2}({\bf x};t)
& = &  
q({\bf x};t) 
+ \int d{\bf y} \; 
\Xi({\bf x}, {\bf y}; \lambda_{0}) q({\bf y};t).
\label{eq:integal_eq_solved}
\end{eqnarray}
The function $\Xi({\bf x}, {\bf y}; \lambda)$ is called the resolvent. 
$\lambda_{0}$ is chosen to make the series convergent. 
For large $N$, the kernel $K({\bf x},{\bf y})$ is $\sim 1/N$. 
Thus, for large $N$, we can take as $\lambda_{0}=1$.
Therefore, if the kernel is bounded, for large $N$, 
we get a convergent series. 
The resolvent is given by 
\begin{eqnarray}
\Xi({\bf x}, {\bf y}; \lambda) 
= \frac{D({\bf x},{\bf y};\lambda)}{D(\lambda)},
\end{eqnarray}
where
\begin{widetext}
\begin{eqnarray}
D(\lambda) & = & 
1 - \lambda \int d{\bf s} \; K({\bf s},{\bf s}) + 
\frac{\lambda^{2}}{2!} \int \int d{\bf s}_{1}d{\bf s}_{2} \; 
K\left (
\begin{array}{ll}
{\bf s}_{1} & {\bf s}_{2} \\
{\bf s}_{1} & {\bf s}_{2} 
\end{array} 
\right )
\nonumber \\
& & 
+ \cdots + 
\frac{(-\lambda)^{p}}{p!} \int \cdots \int d{\bf s}_{1}\cdots d{\bf s}_{p}\; 
K\left (
\begin{array}{llll}
{\bf x} & {\bf s}_{1} & \cdots & {\bf s}_{p}\\
{\bf y} & {\bf s}_{2} & \cdots & {\bf s}_{p}\\
\end{array}
\right )
+ \cdots ,
\label{eq:D_lambda}
\end{eqnarray}
and 
\begin{eqnarray}
D({\bf x},{\bf y};\lambda)
& = & 
K({\bf x},{\bf y}) 
- 
\lambda \int d{\bf s} \; 
K\left (
\begin{array}{ll}
{\bf x} & {\bf s} \\
{\bf y} & {\bf s}
\end{array}
\right )
+ \frac{\lambda^{2}}{2!} 
\int \int d{\bf s}_{1}d{\bf s}_{2}\; 
K\left (
\begin{array}{lll}
{\bf x} & {\bf s}_{1} & {\bf s}_{2} \\
{\bf y} & {\bf s}_{2} & {\bf s}_{2}
\end{array}
\right )\nonumber \\
& & 
+ \cdots + 
\frac{(-\lambda)^{p}}{p!} \int \cdots \int 
d{\bf s}_{1}\cdots d{\bf s}_{p}\; 
K\left (
\begin{array}{llll}
{\bf x} & {\bf s}_{1} & \cdots & {\bf s}_{p}\\
{\bf y} & {\bf s}_{2} & \cdots & {\bf s}_{p}
\end{array}
\right )
+ \cdots,
\label{eq:D_x_y_lambda}
\end{eqnarray}
and
\begin{eqnarray}
 K\left ( 
\begin{array}{llll}
{\bf s}_{1} & {\bf s}_{2} & \cdots & {\bf s}_{p}\\
{\bf t}_{1} & {\bf t}_{2} & \cdots & {\bf t}_{p}
\end{array}
\right ) 
& = & 
\left |
\begin{array}{cccc}
K({\bf s}_{1},{\bf t}_{1}) & K({\bf s}_{1},{\bf t}_{2}) 
& \cdots & K({\bf s}_{1},{\bf t}_{p}) \\
K({\bf s}_{2},{\bf t}_{1}) & K({\bf s}_{2},{\bf t}_{2}) 
& \cdots & K({\bf s}_{2},{\bf t}_{p}) \\
\cdots & \cdots & \cdots & \cdots \\
K({\bf s}_{p},{\bf t}_{1}) & K({\bf s}_{p},{\bf t}_{2}) 
& \cdots & K({\bf s}_{p},{\bf t}_{p})
\end{array}
\right |.
\end{eqnarray}
Thus, a derived kinetic equation is 
\begin{equation}
\frac{\partial f_{1}}{\partial t} 
+ \langle {\bf v}_{1} \rangle \cdot \nabla_{1} f_{1} = 
\int d{\bf r}_{2}\; {\cal L}_{12} g_{2}({\bf r}_{1},{\bf r}_{2};t),
\label{eq:Balescu-Lenard}
\end{equation}
where
\begin{eqnarray}
g_{2}({\bf r}_{1},{\bf r}_{2};t)
& = &  
q({\bf r}_{1},{\bf r}_{2};t) + 
\int d{\bf R} \int d{\bf R}' \; 
\Xi(\{ {\bf r}_{1},{\bf r}_{2} \},\{ {\bf R},{\bf R}' \};\lambda_{0}) \, 
q({\bf R},{\bf R}';t).
\end{eqnarray}
\end{widetext}
This kinetic equation for point vortex system is an analogue 
of the Balescu-Lenard equation in plasma physics. 
Unlike the Balescu-Lenard equation for plasmas, 
the collision term obtained does not have the dielectric function, 
but have additional terms compared with the Landau collision term. 
We note an important point, which manifests a difference 
between point vortex systems and plasma systems. 
The kernel has the factor $1/N$, 
which comes from the circulation $\gamma=\Gamma/N$. 
Thus, the resolvent is expanded in terms of the $1/N$-factor. 
The integral is just the Fourier transform. 
Therefore, the integral is order of $1$. 
In eq.~(\ref{eq:integal_eq_solved}), 
the second term of the right hand side is smaller than 
the first term as the order $O(1/N)$. 
Therefore, for large $N$, we obtain 
\begin{equation}
g_{2}({\bf r}_{1},{\bf r}_{2};t) \approx q({\bf r}_{1},{\bf r}_{2};t).  
\label{eq:g2_q_approx}
\end{equation}
With this result, 
the collision term is reduced to the Landau collision term, 
i.e., eq.~(\ref{eq:Landau_coll1}). 
Therefore, for point vortex systems, 
the Balescu-Lenard collision term is reduced 
to the Landau collision term, because of large $N$-effect, 
i.e., the absence of the dielectric function. 
This is a big difference between point vortex systems and 
plasma systems. 
The screening effect does not appear in point vortex systems, 
while the screening effect is important 
for the Balescu-Lenard equation for plasmas. 
On the Balescu-Lenard collision term in plasma physics, 
see \cite{Balescu} for comparison. \\

In the above paragraph, we showed how to solve the integral equation 
eq.~(\ref{eq:integral_eq_to_be_solved}) formally. 
But the required conditions have not yet been checked. 
Let us look at the required conditions. 
First, to show the operator $T$ 
\begin{equation}
T g_{2} = \int d{\bf y}\; K({\bf x},{\bf y}) g_{2}({\bf y}), 
\end{equation}
is bounded, is important. 
The integral equation is symbolically given by 
\begin{equation}
(1 - \lambda T) g_{2} = q. 
\end{equation}
The formal solution is obtained by the expansion.
\begin{equation}
g_{2} = \sum_{n=0}^{\infty} (\lambda T)^{n} q.
\label{eq:expansion}
\end{equation}
The Fredholm theory for integral equation basically uses 
the boundedness of the operator and the expansion, 
i.e., eq.~(\ref{eq:expansion}). 
This expansion is assured by the boundedness of the operator $T$. 
The expansions of eqs.~(\ref{eq:D_lambda}) and (\ref{eq:D_x_y_lambda}) 
are also due to the boundedness of the operator $T$. 
As the second point, unlike usual Fredholm integral equations, 
the integral domain in eq.~(\ref{eq:integral_eq_to_be_solved}) is infinite. 
If the integral domain is finite and the kernel is bounded, 
the boundedness of operators is easily shown. 
For our case, since the integral domain is infinite, 
we should treat the operator $T$ carefully. 
These two points, i.e., the boundedness of the operator $T$ and 
the infinite integral domain, should be checked and be treated 
in a rigorous way. 
However, in this paper, we do not pursue a rigorous discussion.  
These problems are reserved for mathematical physicists.\\ 
%----------------------------------------------------------------
\section{Concluding remark}\label{sec8}
We have derived a kinetic equation for point vortex systems 
in an infinite plane. 
The kinetic equations derived are analogues of the Landau equation 
and the Balescu-Lenard equation. 
Equation (\ref{eq:Landau}) coincides with the result of 
Chavanis\cite{Chavanis,Chavanis_BBGKY}. 
Equation (\ref{eq:Landau}) possesses several interesting properties, 
which were shown in \cite{Chavanis}. 
The Balescu-Lenard equation (\ref{eq:Balescu-Lenard}) is 
a new kinetic equation. 
Furthermore, we have shown that for large $N$, 
the Balescu-Lenard equation, i.e., eq.~(\ref{eq:Balescu-Lenard}) 
is reduced to the Landau equation, 
i.e., eq.~(\ref{eq:Landau}). 
Therefore, we can conclude that for point vortex systems in an infinite plane, 
without symmetrical restriction 
(such as axisymmetric and unidirectional flows), 
the most generalized kinetic equation is eq.~(\ref{eq:Landau}). 

%----------------------------------------------------------
The following point would be the interesting point 
of the derived kinetic equation, i.e., eq.~(\ref{eq:Landau}):
The interaction among point vortices is long range, 
i.e., logarithmic. 
In addition, the derived kinetic equation is analogous to 
the Landau equation for 3D plasmas. 
However, the collision term for point vortex systems may not diverge. 
In \cite{Chavanis_BBGKY}, for the axisymmetric case, 
the collision term does not diverge. 
This is a symptom of the non-divergence of the collision term 
for the Landau equation. 
The reason of this is that the difference 
between the integration of the collision term, 
i.e., eq.~(\ref{eq:Landau_coll1}), 
and that of the Landau collision term for 3D plasmas. 
The former has the integration with respect to ${\bf r}_{2}$ (i.e., position), 
while the latter has the integration with respect to 
${\bf r}_{2}$ and ${\bf v}_{2}$ (i.e., velocity). 

For the point vortex systems, 
the expression of the energy spectrum was derived 
for the system in an infinite plane\cite{Novikov} and 
for the system in a circular domain\cite{YS}. 
The energy spectrum is closely related to the diffusion coefficient\cite{TM}. 
We will be able to compare the theory of \cite{TM} with 
the kinetic theory in this paper. 

Another interesting point is the following: 
our kinetic theory is not directly connected 
to Onsager's temperature. 
As shown in \cite{YKTSYE} for two-sign point vortex systems, 
Onsager's temperature affects nonequilibrium properties, 
i.e., decaying process. 
To find this connection leads to understand nonequilibirum properties, 
i.e., classification of nonequilibrium processes. 

The most interesting problem, which we would like to attack 
with eq.~(\ref{eq:Landau}), is 
a decaying property of vortex crystals in non-neutral plasmas. 
Unfortunately, eq.~(\ref{eq:Landau}) is 
for the system in an infinite plane, not for the system 
in a circular domain. Thus, it is not for an experimental situation. 
But, eq.~(\ref{eq:Landau}) surely captures the nature of 
phenomena for the system in a circular domain in some extent. 
Many experimental results show that 
the vortex crystals are quasi-stationary states. 
To analyze quasi-stationary states, 
recent advances for long-range interaction systems 
would be some hints for us, 
such as a study of the Hamiltonian mean field (HMF) model.  
The HMF exhibits a slow decay, in which the state is 
stuck in a quasi-stationary state, as well as in point vortex systems. 
For the HMF model, the slow decay is analyzed by the Vlasov equation. 
The quasi-stationary state is very near to a stable solution 
of the Vlasov equation. 
The estimate of some quantities, i.e., the algebraic decay and 
the tail of the velocity distribution function etc., 
is carried out\cite{BandD}, and is tested by a numerical simulation\cite{YBD}. 
Their analysis would be useful for our problems. 
However, to this end, we should know the bahavior of the 2D Euler equation, 
instead that of the Vlasov equation for ususal kinetic theory 
of particle systems. 

%-------------------------------------------------------------------------
\begin{acknowledgments}
The author thanks Professor H.~Tomita for continuous encouragements  
and Professor Y.~Kiwamoto for introducing him vortex dynamics. 
The author is grateful to one of anonymous referees 
for numerous advices to the first manuscript 
and for letting the author know the reference \cite{Chavanis_BBGKY}. 
Some important changes have been made between the first version 
and the second one using the result of \cite{Chavanis_BBGKY}. 
\end{acknowledgments}

\end{document}